\documentclass[aps,prb,preprint,groupedaddress]{revtex4}

\usepackage{graphicx}
\usepackage{amsmath}
\usepackage{amsfonts}
\usepackage{amssymb}
\usepackage[usenames,dvipsnames]{color}

\begin{document}


\title{Polarization switching and nonreciprocity in symmetric and asymmetric magnetophotonic multilayers with nonlinear defect}


\author{Vladimir R. Tuz}
\email[]{tvr@rian.kharkov.ua}
\affiliation{Institute of Radioastronomy of National Academy of
Sciences of Ukraine, 4, Krasnoznamennaya st., Kharkiv 61002,
Ukraine}
\affiliation{School of Radio Physics, Karazin Kharkiv
National University, 4, Svobody Square, Kharkiv 61077, Ukraine}

\author{Sergei V. Zhukovsky}
\email[]{szhukov@physics.utoronto.ca}
\affiliation{Department of Physics University of Toronto, 60 St.~George Street,
Toronto, ON, M5S 1A7, Canada}

\author{Sergey L. Prosvirnin}
\email[]{prosvirn@rian.kharkov.ua}
\homepage[]{http://ri.kharkov.ua/prosvirn/}
\affiliation{Institute of Radioastronomy of National Academy of
Sciences of Ukraine, 4, Krasnoznamennaya st., Kharkiv 61002,
Ukraine} \affiliation{School of Radio Physics, Karazin Kharkiv
National University, 4, Svobody Square, Kharkiv 61077, Ukraine}



\begin{abstract}
A one-dimensional magnetophotonic crystal with a nonlinear defect
placed either symmetrically or asymmetrically inside the structure
is considered. Simultaneous effects of time-reversal nonreciprocity
and nonlinear spatial asymmetry in the structure are studied.
Bistable response is demonstrated in a such system, accompanied by
abrupt polarization switching between two circular or elliptical
polarizations for transmitted and reflected waves. The effect is
explained in terms of field localization at defect-mode spectral
resonances and can be used in the design of thin-film optical
isolators and polarization transformation devices.
\end{abstract}




\maketitle


\newcommand{\sergei}[1]{ \textcolor{Turquoise}{({\bf Comment from Sergei:} #1)}}

\section{Introduction}

Magnetophotonic crystals (MPC's) are periodic structures that
contain magnetic materials and have a period comparable to the
wavelength of electromagnetic radiation
\cite{inoue_jap,sakaguchi,lyubchanskii_rev,zvezdin,inoue_rev,chernovtsev}.
The simplest example of such a periodic structure is a multilayer
having one-dimensional (1D) periodicity. The main advantage of MPC's
in contrast to conventional nonmagnetic photonic crystals (PC's) is
their possibility to tune the band edge position in the spectrum of
the electromagnetic radiation by means of an external static
magnetic field. Moreover, the geometric structure of MPC's allows to
obtain strong enhancement in a number of magneto-optical effects.

Among the magneto-optical effects that can be significantly
enhanced in MPC's, two phenomena are of great interest: (i) the Faraday
effect, which denotes rotation of the polarization ellipse of
light as it propagates collinearly with an externally applied static
magnetic field, and (ii) the nonreciprocity effect, which involves a
difference in phase retardation, polarization rotation, and absorption of
forward- vs.~backward-directed waves propagating through the system.

The Faraday effect can be seen as the lifting of degeneracy for the
left (LCP) and right (RCP) circular polarization states, causing the
LCP and RCP components to propagate with different phase velocities
in the magnetic medium. This difference in velocity of propagation
causes the polarization ellipse of the light to rotate as the light
propagates. The effect is linear with respect to the static magnetic
field strength. The enhancement of this rotation in MPC's originates
from localization of light provided by the multiple interference.
\cite{lyubchanskii_rev,inoue_rev} In fact, the total rotation angle
becomes greater in MPC's with microcavity structure where a magnetic
defect is introduced into the periodic system\cite{inoue_jap}.

Optical nonreciprocity refers to different properties of a medium
for electromagnetic waves propagating in opposite directions. It is
well known that the nonreciprocity effects are inherent to magnetic
media and it can be explained from the symmetry viewpoint.
\cite{kotov} Magnetic field, which is an axial vector, has the
symmetry of circular currents set out in a plane perpendicular to
its vector. For a medium placed in magnetic field, the rotation
directions in this perpendicular plane are non-equivalent. Therefore
the optical properties of a magnetic medium are described with a
non-symmetric permittivity tensor, and the equations for propagation
of LCP and RCP waves in the direction of the field have to be
different. Thus, circularly polarized waves of opposite handedness
(or traveling in opposite directions) are characterized by different
phase velocities and/or attenuations in the course of traveling
along the same optical path. Since the transformation of a
forward-propagating wave into a backward-propagating one with the
same handedness is given by the time-reversal operation, the
sensitivity of medium properties to the reversal of wave propagation
direction is commonly viewed  as the time-reversal nonreciprocity.

Aside from the asymmetry of the permittivity tensor, the material
nonlinearity can be another source of apparent reciprocity failure.
As an example, a nonreciprocal response appears in a layered medium
in which frequency changing or self-focusing is asymmetrically
located and in which there is also nonuniform
dichroism\cite{potton}. The order in which the nonlinear and
dichroic layers are encountered by incident light will significantly
influence the balance between nonlinear and absorptive effects.
Another possible way to obtain nonreciprocal response is to combine
the nonlinearity in a PC with an asymmetrically arranged defect
(i.e., microcavity) inside it, containing some intensity-dependent
material (e.g. a Kerr-type medium). In this system the strong field
localization inside the defect can be achieved, and the internal
field intensity becomes sufficient to change the optical
characteristics of the microcavity through the Kerr effect. Since
the spatial  field distribution is different for the waves incident
on a spatially asymmetric structure from opposite sides,
nonreciprocal response appears\cite{lidorikis,diao}. It is
convenient to call such kind of spatially asymmetric response the
reversible nonreciprocity, since no time-reversal symmetry breaking
takes place here.

It is also important to note that such nonlinear reversible
nonreciprocity is accompanied by optical bistability. Thus, strong
field localization in a defect within a PC alters the
electromagnetic radiation spectrum including the position of the
band edges. This dynamical band edge shift produces optical
bistability which consists in the existence of two stable
transmission or reflection states for the same input intensity; the
typical input-output characteristic of the system contains a
hysteresis loop\cite{grigoriev,smirnov}. In this case the
nonreciprocity manifests itself in the different intensity level of
input light sufficient to achieve bistable switching for the waves
impinging on the system from the opposite sides.

One of the prominent applications of reversible nonreciprocity is
the design of a  nonlinear electromagnetic
diode\cite{hou,grigoriev,miroshnichenko,smirnov}. On analogy with an
electronic diode that transmits electric current in only one
direction due to its nonlinear current--voltage characteristics, the
nonlinear optical diode features unidirectional transmission of the
incoming light. By introducing nonlinearity into the MPC, such
unidirectional transmission can be achieved for one circular
polarization while remaining transparent for the polarization of
opposite handedness.

Hence it is of special interest  to study the simultaneous effects
of time-reversal nonreciprocity and nonlinear spatial asymmetry on
the optical properties of PC's. In this paper, we consider an MPC
where a nonlinear defect which is placed either symmetrically or
asymmetrically inside the periodic structure. An important feature
of the studied system is the fact that the asymmetric bistable
transmission is accompanied by the polarization
conversion\cite{flytzanis,jonson,lyubchanskii}. The main objective
of our study is focused on achieving the bistability-induced abrupt
switching between two distinct polarization states. This can be
important for thin-film polarization optics devices and
polarization-sensitive integrated optics.

The rest of the paper is organized as follows. In
Section~\ref{sec:theory}, we formulate the problem under study and
introduce its solution based on the transfer matrix method of
multilayer optics. Sections~\ref{sec:res1} and \ref{sec:res2} follow
with the results for a nonlinear defect placed symmetrically and
asymmetrically into an MPC, respectively. Finally,
Section~\ref{sec:conclus} summarizes the paper.

\section{Problem formulation and solution\label{sec:theory}}

\begin{figure}[htb]
\centerline{\includegraphics[width=10.0cm]{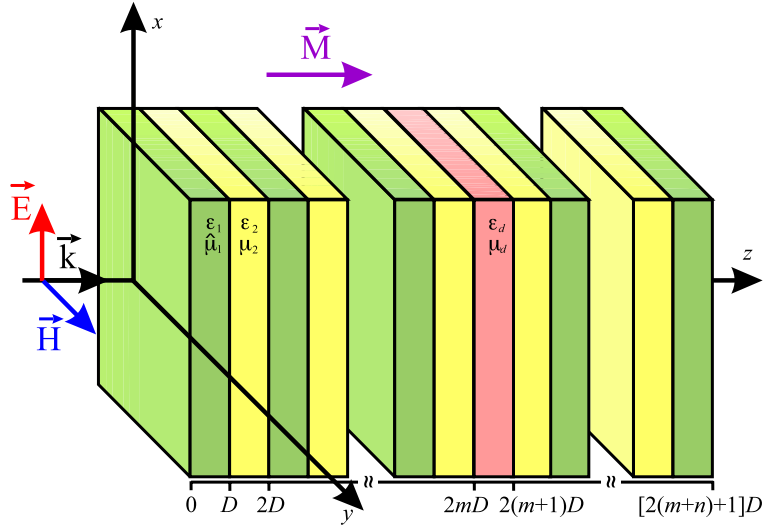}}
\caption{(Color online) Magnetophotonic structure with nonlinear
defect.} \label{fig:fig1}
\end{figure}

We consider a planar multilayer stack of infinite transverse extent
(Fig.~\ref{fig:fig1}). Each unit cell is composed of a bilayer which
consists of magnetic (with constitutive parameters $\varepsilon_1,
\hat \mu_1$) and nonmagnetic (with parameters $\varepsilon_2,
\mu_2$) layers. The magnetic layers are magnetized up to saturation
by an external static magnetic field $\vec M$ directed along the
$z$-axis (Faraday configuration). A defect is created by introducing
into the structure a layer with constitutive parameters
$\varepsilon_d$, $\mu_d$. We assume that this layer is a Kerr
nonlinear dielectric, which permittivity $\varepsilon_d$ linearly
depends on the intensity $|E|^2$ of the electric field. The defect
can be settled either symmetrically or asymmetrically in the middle
of the structure. The parameters $m$ and $n$ describe the number of
bilayers placed before and after the defect layer. In any case
the bilayers are arranged symmetrically with respect to the
defect layer, i.e. the structure begins and ends with layers of
the same type. We suppose that all layers have the same thickness
$D$. The outer half-spaces $z\le 0$ and $z\ge [2(m+n)+1]D$ are
homogeneous, isotropic, and have parameters $\varepsilon_0, \mu_0$.
Assume that the normally incident field is a linearly polarized
plane monochromatic wave of a frequency $\omega$ and an amplitude
$A$. For the sake of definiteness, we also suppose that the vector
$\vec E$ of the incident wave is directed along the $x$-axis.

As an convenient material for magnetic layers, the family of
impurity-doped yttrium-iron garnet (YIG) Y$_3$Fe$_5$O$_{12}$ films
can be proposed. These magnetic oxides are well studied and widely
used in integrated magneto-optics because they are transparent in
the near infrared region\cite{lyubchanskii_rev,inoue_rev}. As an
example, a few types of multilayered films composed of magnetic
Bi-substituted YIG (Bi:YIG) and dielectric SiO$_2$ or glass FR-5
layers were investigated. \cite{sakaguchi,kotov} The MPC's based on
the other materials are also known. Thus, a new class of
semiconductor-magnetic hybrid nanostructures consist of GaAs with
MnAs nanoclasters (GaAs:MnAs) which are paired with GaAs/AlAs
superlatices is recently investigated experimentally in the range
900-1100~nm \cite{shimizu}. Also in the nonlinear regime the
structure based on the semimagnetic semiconductors such as
Cd$_{1-x}$Mn$_x$Te with the defect being a quantum well with
prescribed spectral characteristics was reported\cite{buss,cubian}.
From these papers it may be deduced that the magnetic materials
manifest their nonlinear properties at the light intensity about
1~GW/cm$^2$. In our present paper we consider the nonlinear defect
which is made of nonmagnetic material due to its greater
availability. As an example, AsGa or InSb can be selected for this
purpose. We prefer such structure configuration because these
materials require much lower intensities of the incident light to
enable the nonlinear effects. From the literature \cite{palik} it
can be deduced that the nonlinear response in the semiconductor
materials can be achieved at the light intensity about 1~kW/cm$^2$.
Although a defect is made of nonmagnetic material, the studied
structure that consists of magnetic layers and such nonlinear defect
exhibits a number of very interesting and unique properties that we
consider.

Our solution is based on the transfer matrix
formalism\cite{berreman} which is used to calculate the field
distribution inside the structure and the reflection and
transmission coefficients of the MPC. In the Faraday configuration,
when external static magnetic field is biased parallel to the
direction of wave propagation ($\vec k\parallel \vec M$), the
magnetic permeability $\hat \mu_1$ is a tensor quantity with nonzero
off-diagonal components:
$$\hat \mu_1=
\begin{pmatrix}
\mu_1^T & i\alpha   & 0 \\
-i\alpha  & \mu_1^T & 0 \\
0         & 0       & \mu_1^L
\end{pmatrix}.
$$

For the description of electromagnetic waves in this case it is
necessary to use a $4 \times 4$ transfer matrix
formulation\cite{vidil}. Thus, at the first stage, in the linear
case, the equation which defines the coupling of the tangential
field components at the input and output of the structure is written
in the next form\cite{tuz_josab_09a,tuz_josab_11}
\begin{equation}\label{eq:one}
\vec{\Psi} (0)=\mathfrak{M} \vec{\Psi}(\Lambda)=\left\{(
\mathbf{M}_1 \mathbf{M}_2)^m \mathbf{M}_d(\mathbf{M}_2
\mathbf{M}_1)^n\right\}\vec{\Psi}(\Lambda),
\end{equation}
where $\vec{\Psi}=\{ E_x,E_y,H_x,H_y\}^T$ is the vector containing
the tangential field components at the structure input and output;
the upper index $T$ is the matrix transpose operator; $\Lambda$ is
the total length of the structure, $\Lambda=[2(m + n) + 1]D$; $m$
and $n$ are the numbers of periods placed before and after the
defect element; $\mathbf{M}_1$, $\mathbf{M}_2$ and $\mathbf{M}_d$
are the transfer matrices of the rank four of the first, second, and
defect layers, respectively. The elements of the transfer matrices
in (\ref{eq:one}) are determined from the solution of the Cauchy
problem and are given in\cite{vidil}.

As the solution of the linear problem (\ref{eq:one}) is obtained,
the intensity of the reflected and transmitted fields and the
distribution of the field $\vec {E}_{in}(z)$ inside the MPC can be
calculated. Generally, when the defect layer consists of a Kerr
nonlinear dielectric, the permittivity $\varepsilon_d$ is
inhomogeneous, and depends on the intensity of the electric field at
each point of this layer as follows
\begin{equation}
\varepsilon_d(z)=\varepsilon_d^l+\varepsilon_d^{nl}
|E_{in}(z)|^2,~(2mD \le z \le (2m+1)D). \label{eq:two}
\end{equation}
Knowing the field intensity in the defect layer, both the actual
value of permittivity $\varepsilon_d$ and, consequently, the actual
value of transfer-matrix $\mathfrak{M}$ can be calculated.  Thus we
deal with an equation on the unknown function of field intensity
distribution inside the defect layer. A magnitude of the incident
field $A$ is an independent parameter of this equation. Since the
parameter $\varepsilon_d^{nl}$ is small and the nonlinear
contribution to $\varepsilon_d$  varies with the longitudinal
distance on the scale of one-half wavelength we provide an approach
which regards $\varepsilon_d$ as independent on $z$ and treats the
dependence of $\varepsilon_d$ on the average intensity of the
electric field $\overline{|E_{in}|^2}$ inside the defect layer.
Quantitative reasoning of this approach is presented
in\cite{tuz_josab_11}. On the basis of this approximation, we
suppose that the permittivity of the medium depends on the average
intensity of the electric field as $\varepsilon_d=\varepsilon_d^l+
\varepsilon_d^{nl} |\overline{E_{in}|^2}$.

As a result, at the second stage, the nonlinear equation related to
the average field intensity distribution in the defect is obtained.
The numerical solution of this equation yields us the final field
distribution in the MPC and the values of the reflection $R$ and
transmission $T$ coefficients, which expressions can be found
in\cite{vidil}.

\section{Symmetric multilayers: Polarization bistability\label{sec:res1}}
Our objective here is to study the main features of optical
response for an MPC with a nonlinear defect placed  symmetrically inside
it. For this reason we consider an MPC consisting of two sections
with the same number of bilayers in them ($m=n$). The sections are
located symmetrically on each side of the defect layer. The main
idea of such an arrangement is to obtain a significant field
localization inside the defect layer, which is achieved by
an appropriate choice of the  number of periods $m$ and the
material parameters of layers.

The basic optical properties of the studied MPC are inherited from
the characteristics of perfectly periodic structures with
nonmagnetic layers. Recall that all periodic structures with layer
thicknesses comparable to the wavelength possess forbidden frequency
gaps (stopbands or band gaps) as a direct consequence of
Floquet--Bloch theorem.\cite{sakoda} These gaps are determined by
the modulation period and the average refractive index. Propagation
of waves with frequencies in the stopbands of an idealized infinite
structure is completely inhibited, and the band gaps are in this
sense perfect. For finite structures these gaps appear as frequency
regions with low transmittance and high reflectance, located between
high-transmittance passbands. If any distortion (a ``defect'') is
introduced inside a periodic structure, transmission resonances can
appear in the stopbands, with the field strongly localized in the
defect. The existence of such ``localization resonances'' is
explained by the fact that the defect forms a resonant Fabry--Perot
cavity enclosed between two Bragg mirrors.

\begin{figure}[htb]
\centerline{\includegraphics[width=10.0cm]{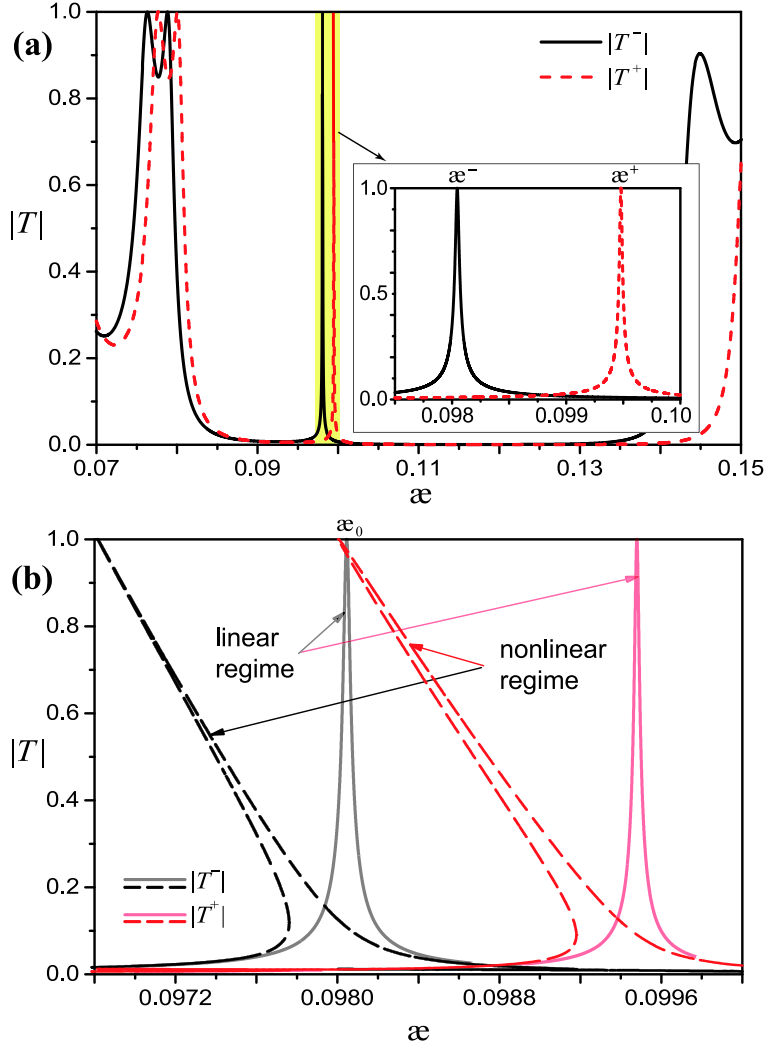}}
\caption{(Color online) Frequency dependences
($\varkappa=D/\lambda$) of the transmission coefficient ($T$) of the
LCP ($-$) and RCP ($+$) waves in the (a) linear  (b) nonlinear  case
for $m=n=5$, $\varepsilon_1=10$, $\mu_1^T=\mu_1^L=1$, $\alpha=0.05$,
$\varepsilon_2=\mu_2=\mu_d=1$, $\varepsilon_d^l=4$. For the
nonlinear case,
$\tilde\varepsilon_d^{nl}=\varepsilon_d^{nl}I_0=1.5\times 10^{-4}$,
which corresponds to the incident light intensity
$I_0=15\,\text{kW}/\text{cm}^2$ for
$\varepsilon_d^{nl}\simeq1.0\times10^{-5}\,\text{cm}^2/\text{kW}$.
} \label{fig:fig2}
\end{figure}
\begin{figure}[htb]
\centerline{\includegraphics[width=10.0cm]{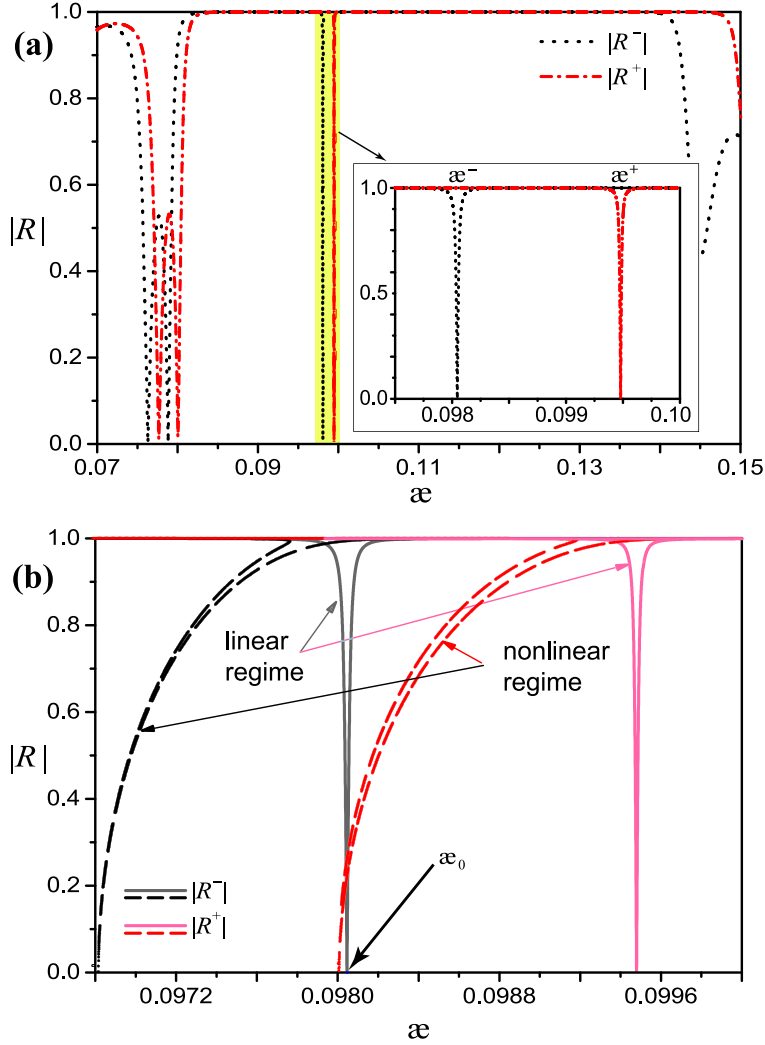}}
\caption{(Color online) Same as Fig.~\ref{fig:fig2} for the
reflection coefficient ($R$).} \label{fig:fig3}
\end{figure}

The main distinctive feature of an MPC in contrast to the
nonmagnetic one is the appearance of circular polarization
eigenstates. Such circular polarization eigenstates are also
inherent in PC's with chiral isotropic
layers\cite{tuz_josab_09a,tuz_josab_11} but in the case of MPC's
they are controlled with an external static magnetic field. Thus the
MPC reacts differently to circularly polarized waves with opposite
handedness, with distinct optical spectra for each of them (see
Figs.~\ref{fig:fig2}a and \ref{fig:fig3}a). This way, in the Faraday
configuration, both the edges of the forbidden bands and the
frequencies of the localized defect modes become different for LCP
vs.~RCP incident wave. As a result, the defect resonances split into
doublets (see Figs.~\ref{fig:fig2}a--\ref{fig:fig3}) known as the
longitudinal Zeeman-like doublets\cite{jonson}. These doublets
originate from lifting of the degeneracy between resonant conditions
for the LCP and RCP waves in the underlying MPC by the external
magnetic field. It can be seen that there are two closely spaced
resonant modes in the stopband, one of which is an RCP eigenmode and
the other is an LCP eigenmode.

In the insets of Figs.~\ref{fig:fig2}--\ref{fig:fig3} the frequency
band where the doublet exists is given on a larger scale. Throughout
the paper we suppose that the working frequency is far from the
frequency of the ferromagnetic resonance of magnetic layers and
their losses are negligibly small. Under this assumption, at the
resonant frequencies, the magnitude of the transmission coefficient
of the corresponding circularly polarized mode reaches unity, and
the structure becomes completely transparent for the LCP wave when
$\varkappa^-\approx0.098$ and for the RCP wave when
$\varkappa^+\approx0.0995$. Obviously, the magnitude of splitting
(the frequency difference between the peaks $\Delta \varkappa =
\varkappa^+ - \varkappa^-$) can be easily tuned by changing the
strength of the external static magnetic field.

Now we consider the case when the MPC contains a Kerr-type nonlinear
defect. It is known that the introduction of such a defect into an
otherwise linear structure can induce bistable behavior in the
system. The nature of this bistability is studied in the theory of
the nonlinear Fabry--Perot resonators quite well.\cite{gibbs} The
resonant frequencies $\varkappa^\pm$ are sensitive to the refractive
index of the material within the cavity. Thus, when the frequency of
the incident wave is tuned near a resonant frequency, the field
localization induces growth of the light intensity inside the
cavity, which, by means of the Kerr effect, eventually alters the
refractive index enough to shift the resonant frequency. When this
shift brings the resonant condition closer to match the frequency of
the incident field, even more energy gets localized in the cavity.
This further enhances the shift of the resonance, creating positive
feedback that leads to formation of a hysteresis loop in the spectra
with respect to the incident field intensity. As a result, for a
fixed input field intensity, the frequency dependences for any
resonant mode have a typical shape of ``bent resonances''. In the
spectra of a nonlinear MPC this bending can be seen for both
resonant modes in the split doublets
(Figs.~\ref{fig:fig2}b--\ref{fig:fig3}b).

\begin{figure}[htb]
\centerline{\includegraphics[width=10.0cm]{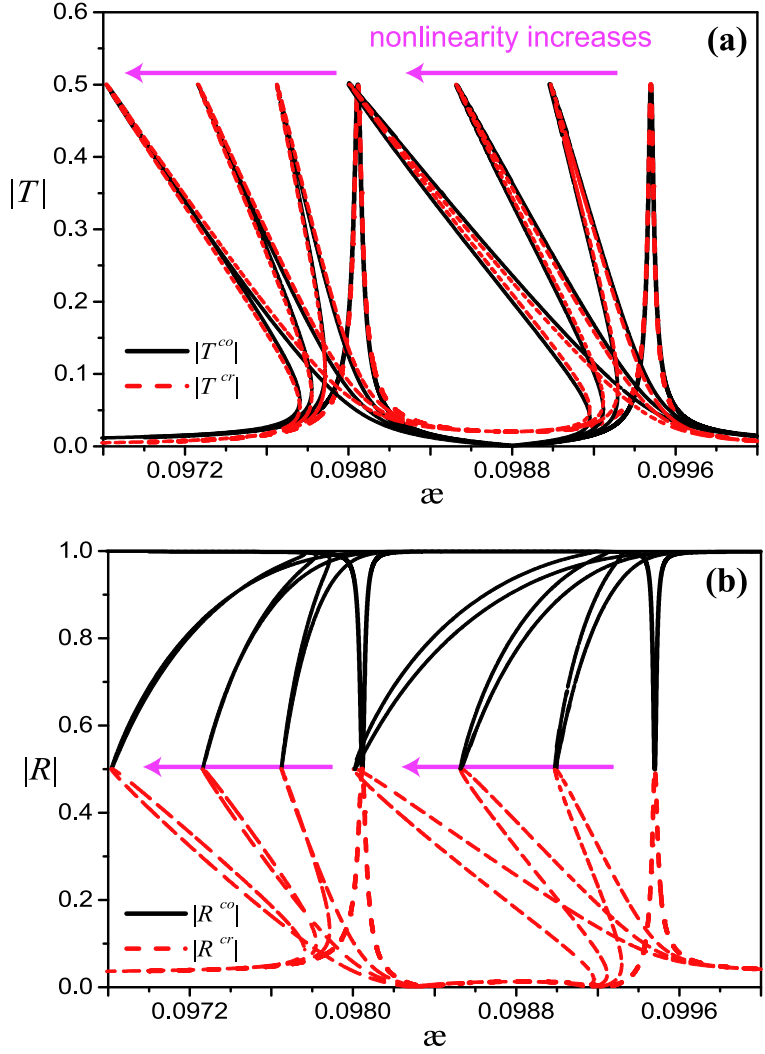}}
\caption{(Color online) Frequency dependences
($\varkappa=D/\lambda$) of the magnitudes of the co-polarized (co)
and cross-polarized (cr) components of the transmission (a) and
reflection (b) coefficients of linearly polarized waves. The input
intensity $I_0$ in the nonlinear regime is taken to be 5, 10, and 15
$\text{kW}/\text{cm}^2$. Other parameters are as in
Fig.~\ref{fig:fig2}.} \label{fig:fig4}
\end{figure}

Now consider a linearly polarized wave incident on an MPC with
defect. One can represent it as a superposition LCP and RCP waves.
As a result, the corresponding optical spectra will contain both
resonances. This is demonstrated in Fig.~\ref{fig:fig4} for
individual polarization components of reflected and transmitted
light, as measured in typical experiments. Since the whole system
possesses axial symmetry in the considered case of normal incidence
and Faraday configuration, we can only distinguish between
co-polarized (e.g., $ss$ or $pp$, denoted $co$) and cross-polarized
($sp$ or $ps$, denoted $cr$) components. Since LCP and RCP cmponents
are present in a linearly polarized wave in equal proportion, the
magnitudes of the co-polarized and cross-polarized components are
equal to each other at the resonant frequencies,
$|T^{co}|=|T^{cr}|=|R^{co}|=|R^{cr}|=0.5$. These conditions are
satisfied in the both linear and nonlinear regimes. In the nonlinear
case, both localization resonances are bent. The ``angle'' of
bending clearly depends on the intensity of the incident field and
is almost the same for both resonances in the doublet.

\begin{figure}[htb]
\centerline{\includegraphics[width=12.0cm]{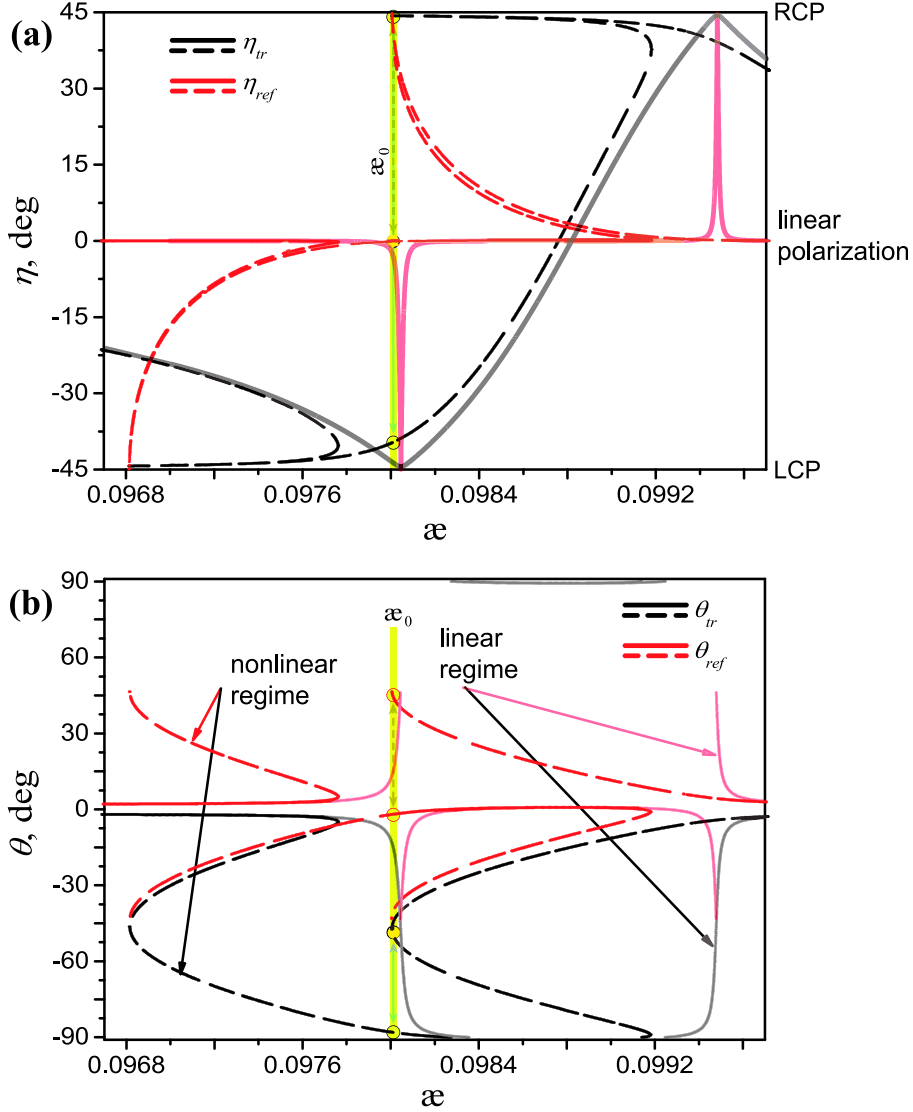}}
\caption{(Color online) Frequency dependences
($\varkappa=D/\lambda$) of
 (a) the elipticity angle $\eta$ and (b) the polarization azimuth $\theta$ of the
transmitted and reflected fields. The incident light is linearly
polarized, and structure parameters are as in Fig.~\ref{fig:fig2}.
The vertical line marks the bistable polarization switching at
$\varkappa_0$.} \label{fig:fig5}
\end{figure}

Due to the above mentioned polarization sensitivity of a
magnetophotonic system, a linearly polarized wave will very likely
undergo a change in its polarization state during reflection or
transmission.  This is confirmed in Fig.~\ref{fig:fig5}, which shows
the corresponding frequency dependences of the ellipticity angle
($\eta$) and the polarization azimuth ($\theta$)  for the
transmitted (black lines) and reflected (red lines) fields.
According to the definition of the Stokes parameters, we introduce
the ellipticity $\eta$ so that the field is linearly polarized when
$\eta=0$, and $\eta=-\pi/4$ for LCP and $+\pi/4$ for RCP (note that
in the latter cases the preferential azimuthal angle of the
polarization ellipse $\theta$ becomes undefined). In all other cases
($0<|\eta|<\pi/4$), the field is elliptically polarized. In the
considered frequency band and in the linear regime, the transmitted
field experiences the rotation of its polarization ellipse and
sequentially changes between LCP and RCP through elliptical and
linear polarization states (Fig.~\ref{fig:fig4}, solid black lines).
On the contrary, the reflected field is linearly polarized almost in
the whole selected band except the frequencies $\varkappa^-$ and
$\varkappa^+$ where it becomes circularly polarized
(Fig.~\ref{fig:fig5}, solid red lines). Note that at these resonant
frequencies the polarization azimuth
$\theta_{ref}=\theta_{ref}(\varkappa)$ is a discontinuous function.

Such a drastic difference in the polarization states of the
transmitted vs.~reflected fields can be understood from the fact
that the operating frequencies lie in the stopband of the MPC where
an impinging wave is almost completely reflected from the structure.
As the incident field is linearly polarized, so, too, is the
reflected field. Due to the finite size of the structure a small
fraction of the wave's energy still gets transmitted through the
MPC, undergoing a $90^\circ$ rotation of its polarization ellipse
(Fig.~\ref{fig:fig5}b) for $\varkappa^-<\varkappa<\varkappa^+$. At
the resonant frequencies, it is evident that the matching circularly
polarized eigenmode passes through the system while for the
orthogonally polarized eigenmode the transmission is still
forbidden. Therefore, both transmitted and reflected fields become
circularly polarized within the localized modes frequencies. Note
that the reflected field has the same polarization state as the
transmitted field because the reflected wave propagates in the
opposite direction (see Ref.~\onlinecite{kotov} for clarity).

In the nonlinear regime the ellipticity angle and the polarization
azimuth become multivalued functions. Therefore, it is possible to
use multistability to switch not only between different
transmittances and reflectances but also between two (or, generally,
more than two) distinct polarization states in the transmitted
and/or reflected light.

The most intriguing scenario for such switching is expected when a
bent resonance at $\varkappa^+$ spectrally overlaps with the
original location of the other resonance at $\varkappa^-$. This
overlap is possible as the resonances are spectrally close to each
other. For example, let us fix the operating frequency $\varkappa_0$
at $\varkappa^-$. At this frequency the reflected and transmitted
fields ought to be LCP. As the intensity of input field rises, the
other resonance corresponding to $\varkappa^+$ and associated with
RCP undergoes red shift and eventually reaches $\varkappa_0$.
It becomes possible to couple the incident wave with frequency
$\varkappa_0$ with either of the eigenmodes. Since these have
opposite circular polarizations (they are associated with converting
a linearly polarized incident light into LCP and RCP), it can be
expected that switching between these two polarization states can be
achieved.


Indeed, Fig.~\ref{fig:fig5} shows that at a frequency
$\varkappa_0\approx \varkappa^{-}$ the bistable switching occurs
between RCP and near-LCP for the transmitted light and between
linear polarization and RCP for the reflected light. This agrees
with the above explanation and is seen in the behaviour of resonance
bending in the Stokes parameter space (Fig.~\ref{fig:fig5}). For the
reflected light the bending in ellipticity resembles that in the
reflectance (Fig.~\ref{fig:fig4}a). For the transmitted light the
bent resonances occur in the immediate vicinity of $\eta=\pm \pi/4$,
because only circularly polarized waves can fully couple to the MPC
eigenmodes to become transmitted through it.


Finally, note that Fig.~\ref{fig:fig4}b illustrates another
peculiarity of the reflection spectra of the structure under study,
namely, the formation of closed loops, which appear in the
cross-polarized component of the reflected field. In particular, the
closed loop appears in the lower-frequency resonance at
$\varkappa^-$. The physical mechanism of loop formation is the
difference between the values of $T_{co}$ and $R_{co}$ to either
side of the resonance. In the linear regime,
$|T_{co}(\varkappa^{-}-\delta)|<|T_{co}(\varkappa^{-}+\delta)|$
since transmittance between the resonances should be higher that to
the either side of both defects because it is influenced by the
Lorentzian tails of both resonances. Consequently,
\begin{equation}|R_{co}(\varkappa^{-}-\delta)|>|R_{co}(\varkappa^{-}+\delta)|.\label{eq:inequal}\end{equation}
(This inequality can also be influenced by non-symmetric placement
of the resonances in the band gap due to the violation of the
quarter-wave condition in the structures under study.) In the
nonlinear regime, the relation in Eq.~(\ref{eq:inequal}) holds, and
the resonance bending to the direction of lower frequencies will
cause a loop to form.

\section{Asymmetric configuration: Polarization conversion\label{sec:res2}}

Nonlinear multilayer structures with spatial asymmetry, are commonly considered to obtain directional sensitivity or reversible nonreciprocity  in nonmagnetic PC's.
As a few examples,  random or deterministically
aperiodic media, as well as periodic structures with asymmetrically
positioned defects, were recently reported to have direction-dependent or unidirectional transmission.\cite
{grigoriev,miroshnichenko,smirnov,tuz_josab_11,zhukovsky,bliokh}
The general result is that interaction between nonlinearity and
asymmetry  manifests itself in the simultaneous occurrence of bistability (or multistability) and nonreciprocity.

\begin{figure}[htb]
\centerline{\includegraphics[width=10.0cm]{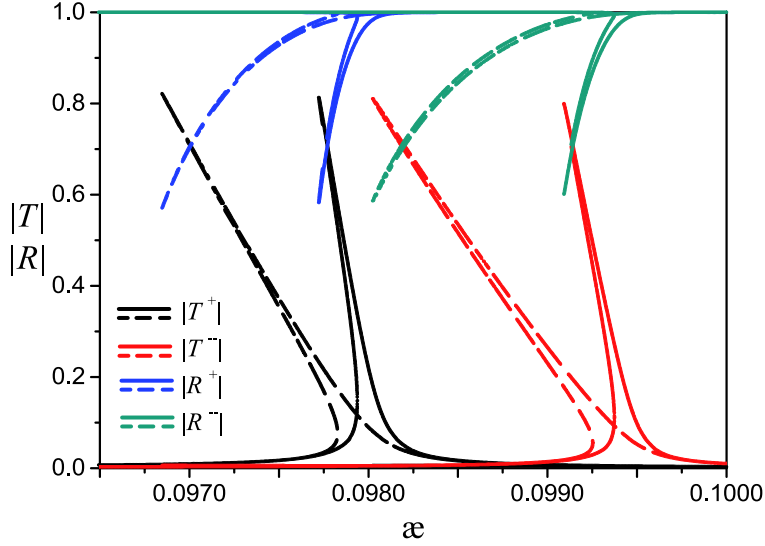}}
\caption{(Color online) Frequency dependences
($\varkappa=D/\lambda$) of the transmission ($T$) and reflection
($R$) coefficients of the LCP ($-$) and RCP ($+$) waves of the MPC
with asymmetrically placed ($m\ne n$) nonlinear defect. Here,
$\tilde\varepsilon_d^{nl}=\varepsilon_d^{nl}I_0=1.0\times 10^{-4}$,
i.e., $I_0=10\,\text{kW}/\text{cm}^2$ for
$\varepsilon_d^{nl}\simeq1.0\times10^{-5}\,\text{cm}^2/\text{kW}$.
Other parameters are as in Fig.~\ref{fig:fig2}. Solid and dash lines
correspond to ($m=5$, $n=6$) and ($m=6$, $n=5$) configurations,
respectively.} \label{fig:fig6}
\end{figure}

From a mathematical point of view, this all-optical reversible nonreciprocity
is a result of non-commutativity of matrix multiplication
in Eq.~(\ref{eq:one}) when the transfer-matrix of
the structure is calculated. In partuicular, optical properties
of a 1D periodic structure with a defect strongly depend on the
position of that defect layer inside the sample. Nevertheless, in
the linear regime,  specific properties of the transfer matrix that
stem from time-reversal reciprocity of the Maxwell equations ensure
that the transmission through the system remains the same regardless
of whether the field is incident from the left or right side of the
structure.

The situation changes drastically if an optically
sensitive (e.g., Kerr-type nonlinear) material is used for the defect
layer. In this case, due to different field localization patterns within
the defect layer for the waves impinging from the left and
right sides of the structure, the nonlinear response becomes different.
This difference manifests itself in the different angles of bending
of the localization resonances.\cite{smirnov}

\begin{figure}[htb]
\centerline{\includegraphics[width=12.0cm]{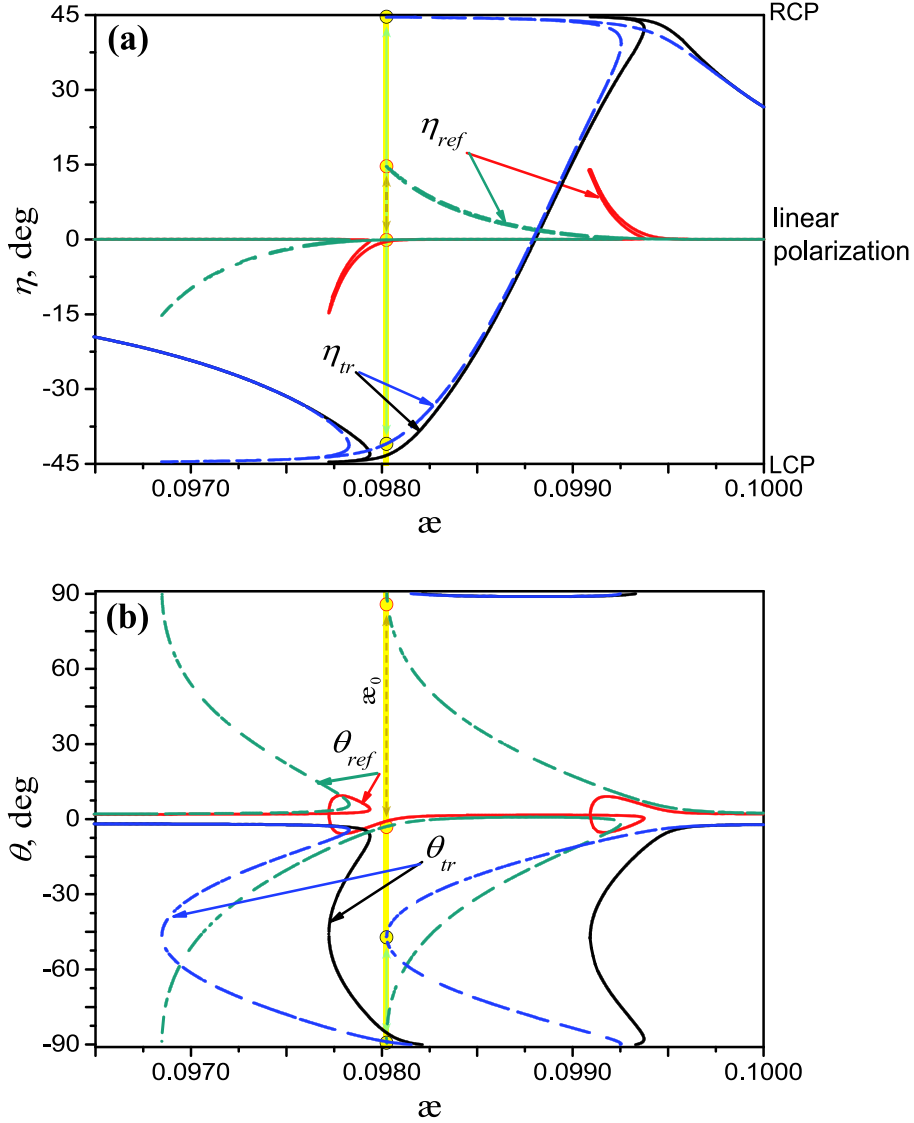}}
\caption{(Color online) Frequency dependences
($\varkappa=D/\lambda$) of the elipticity angle (a) and the
polarization azimuth (b) of the transmitted and reflected fields of
the MPC with asymmetrically placed ($m\ne n$) nonlinear defect.
Parameters are as in Fig.~\ref{fig:fig6}. The vertical line marks
the bistable polarization switching at $\varkappa_0$. }
\label{fig:fig7}
\end{figure}

Our goal here is to study the simultaneous effect of the spatial
asymmetry and the time-reversal nonreciprocity on the behavior of
the localization resonances in the MPC. We modify the structure from
Section~\ref{sec:res1} to make the number of bilayers in two
subsections before and after the defect element different ($m\ne
n$). We additionally assume that the static magnetic field direction
always coincides with the wave propagation direction. This can be
assumed
 without loss of generality because changing the
direction of wave propagation without changing the direction of the
static magnetic field reverses the handedness of  the circularly
polarized states (RCP$\rightleftharpoons$LCP). Hence by considering
the response of the original structure characterized by $(m,n)$ and
its mirror-symmetric counterpart $(n,m)$ to LCP and RCP incident
wave solves the problem completely.

\begin{figure}[htb]
\centerline{\includegraphics[width=10.0cm]{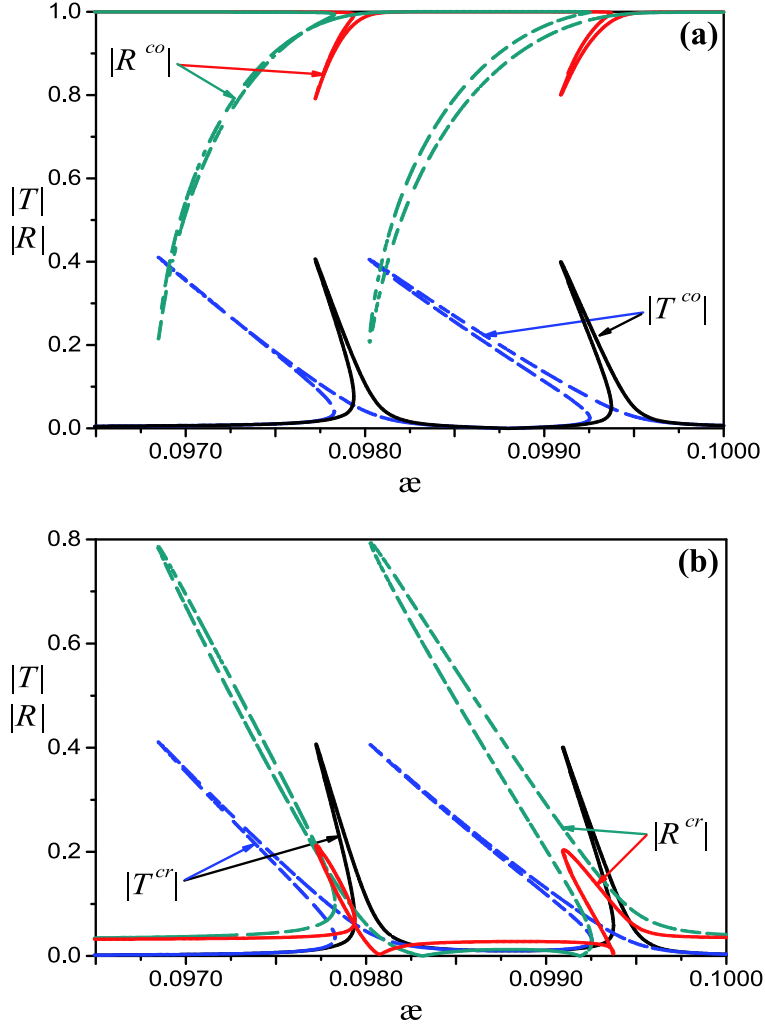}}
\caption{(Color online) Frequency dependences
($\varkappa=D/\lambda$) of the magnitudes of the co-polarized (co)
and cross-polarized (cr) components of the transmission (a) and
reflection (b) coefficients of the linearly polarized waves of the
MPC with asymmetrically placed ($m\ne n$) nonlinear defect.
Parameters are as in Fig.~\ref{fig:fig6}.} \label{fig:fig8}
\end{figure}

Comparison of the results presented in
Figs.~\ref{fig:fig2}b--\ref{fig:fig3}b and Fig.~\ref{fig:fig6} shows
that adding one bilayer at either side of the MPC drastically
changes the spectra of the structure. These changes are associated
with the already mentioned different field distribution inside the
structure.  The stark difference in the angles of the localization
resonance bending results from the all-optical reversible
nonreciprocity.

The accompanying change of the  magnitude for the reflection and
transmission coefficients at the bent resonances (so that
$|T^\pm_\text{max}|<1$ and $|R^\pm_\text{min}|>0$) results from a
certain conflict in the design principles for resonant multilayers.
Namely, to increase the structure's sensitivity to the direction of
incidence, one needs to increase its the spatial asymmetry; yet to
increase the maximum transmission at a resonant peak, the structure
should remain close to symmetric\cite{zhukovsky,smirnov}. As a
consequence, at the frequencies of the localization resonances the
transmission is always below unity and the reflected field is always
elliptically rather than circularly polarized. Indeed, as seen in
Fig.~\ref{fig:fig7}, the ellipticity angle $|\eta_{ref}|<\pi/4$ in
the whole selected frequency band. The transmitted field is still
circularly polarized at the localization resonances. The
polarization azimuth $\theta_{ref}=\theta_{ref}(\varkappa)$ is now a
continuous function.
Hence, while the symmetric structure features  polarization switching  between two
circularly polarized states, the asymmetric one only enables switching between two elliptically polarized states.

However, it can be seen that changing the position of the defect
layer within the structure significantly alters the ratio between
the reflected and transmitted field, and in particular the relations
between co-polarized and cross-polarized components in them
(Fig.~\ref{fig:fig8}). While the magnitudes of the co-polarized and
cross-polarized transmission components remain equal to each other
($|T^{co}|=|T^{cr}|\le0.5$), the the relation between the reflection
components ($|R^{co}|$ and $|R^{cr}|$) varies in a much wider range.
In one structure configuration ($m=5$, $n=6$), the peak magnitudes
of the co-polarized and cross-polarized reflection components are
$|R^{co}|\approx 0.8$ and $|R^{cr}|\approx 0.2$. In the other
configuration ($m=6$, $n=5$) they are opposite: $|R^{co}|\approx
0.2$ and $|R^{cr}|\approx 0.8$. In the latter case there is an
obvious significant polarization transformation in the reflected
field so that a $90^\circ$ polarization rotation of the incident
light can be achieved with good conversion efficiency. This can find
useful application as thin-film tunable polarization-rotating
mirrors. Also, an appropriate choice of the asymmetric structure
configuration, material parameters, layer thicknesses, and magnetic
field strength would achieve switching between two orthogonal linear
polarization states in the reflected field. This can be important in
the design of tunable thin-film polarization splitters and
switchers.

\section{Conclusions\label{sec:conclus}}
In the present paper, we have studied the effects of
bistability, nonreciprocity, and polarization transformation in a
magnetophotonic crystal with a nonlinear defect placed either
symmetrically or asymmetrically inside the structure. The problem is
considered in the Faraday configuration, i.e, the external static
magnetic field is applied in the direction of the structure
periodicity and is collinear with the wave vector of the incident
wave.

The reflection and transmission coefficients of the structures, along with the field distribution inside them, are calculated using the transfer matrix approach. The nonlinear
problem is solved under the assumption that the nonlinear permittivity of the medium inside the defect layer depends on the average intensity of the electric field inside the defect.

In the case of symmetric structure configuration, it is shown that a
bistable response of a nonlinear magnetophotonic system features switching
between two circular polarization states within the localization
resonances (defect modes) for reflected and transmitted fields.
In the case of asymmetric structure configuration, this switching
appears between elliptically polarized states in the reflected
field, and between circularly polarized states in the transmitted field. The asymmetric structure also features strong $90^\circ$ polarization rotation in the reflected field, with a potential for bistable switching between linear polarizations.

From the specific parameters used in our numerical calculations, it is reasonable to conclude
that  bistable response and stepwise polarization switching can already be achieved at the incident power densities of 10--100 kW/cm$^2$ with available materials in the considered structure configuration.

\bibliography{nonlinear_mpc}

\end{document}